\documentclass{article}

\usepackage{arxiv}

\usepackage{float}
\usepackage[utf8]{inputenc} 
\usepackage[T1]{fontenc}    
\usepackage{hyperref}       
\usepackage{url}            
\usepackage{booktabs}       
\usepackage{amsfonts}       
\usepackage{nicefrac}       
\usepackage{microtype}      
\usepackage{lipsum}		
\usepackage{graphicx}
\usepackage{natbib}
\usepackage{doi}

\title{Smoking prevalence in Covid-19 patients}

\author{ 
	\hspace{1mm}Sina Bagheri-Nezhad\\
	School of Computer Engineering\\
	Iran University of Sciecnce and Technology\\
	Tehran, Iran \\
	\texttt{s\_bagherinezhad@comp.iust.ac.ir} \\
	\And
	\hspace{1mm}Nasser Mozayani \\
	School of Computer Engineering\\
	Iran University of Sciecnce and Technology\\
	Tehran, Iran \\
	\texttt{mozayani@iust.ac.ir} \\
	\And
	\hspace{1mm}Elham Abdi \\
	School of Metallurgy and Materials Engineering\\
	Iran University of Sciecnce and Technology\\
	Tehran, Iran \\
	\texttt{e\_abdi@metaleng.iust.ac.ir} \\
	\And
	\hspace{1mm}Setareh Rostami \\
	School of Computer Engineering\\
	Iran University of Sciecnce and Technology\\
	Tehran, Iran \\
	\texttt{rostami\_setareh@vu.iust.ac.ir} \\
}

\renewcommand{\shorttitle}{Smoking prevalence in Covid-19 patients}

\hypersetup{
pdftitle={Smoking prevalence in Covid-19 patients},
pdfsubject={stat.AP},
pdfauthor={Sina Bagheri-Nezhad, Nasser Mozayani, Elham Abdi, Setareh Rostami},
pdfkeywords={smoking, COVID-19, coronavirus, statistical hypothesis testing, runs test, chi-square test},
}

\begin{document}
\maketitle
\begin{abstract}
	We investigate the prevalence rate of smoking in Covid-19 patients and examine whether there is a difference in the distribution of smokers between the two statistical populations of critically ill patients with Covid-19 and the entire Iranian population or not. To do this, we first prepared a sample of 1040 Covid-19 patients admitted to hospitals in Tehran, Rasht, and Bojnord. Then, through the non-parametric statistical runs test, we show that the sample was randomly selected, and it is possible to generalize the result of tests on the sample to the community of hospitalized Covid-19 patients. In continuation, we examined the hypothesis that the smoking prevalence among Covid-19 patients admitted to hospitals is equal to the prevalence rate of smoking in Iranian society. For this purpose, we used the non-parametric chi-square test, and it was observed that this hypothesis is rejected. The data show a significant difference in the prevalence of smoking between critically ill Covid-19 patients and the whole of Iranian society. Additionally, we examined this hypothesis in some subpopulations, and the results were the same.
\end{abstract}


\section{Introduction}
The Covid-19 disease has become a significant public health threat in the last years and has disrupted humans life. During this period, this disease has had a substantial impact on human social life in many ways. In addition to involving the health sector, it has also affected the economic, educational, and cultural systems. For this reason, identifying the factors affecting it is one of the current research questions.

Given that Covid-19 disease is an infectious disease that attacks the lungs \citep{1}, it is assumed that smoking, which is considered a known risk factor for many respiratory infections \citep{1}, causes more severe complications in patients with new coronavirus. Therefore, this study examines this hypothesis.

Indeed, the first hypothesis that comes to mind about the relationship between smoking and Covid-19 disease is that smoking causes more severe complications in patients with this disease. Nevertheless, while researching the factors influencing Covid-19 disease, the authors of this article came across several studies whose results seemed strange. Zhou et al. examined 191 Covid-19 patients in Wuhan, China. They found that only 11 of them smoked \citep{2}. Zheng et al. also reviewed the radiographs of 140 patients with Covid-19 and found that only 7 of them had smoked in the past, and 2 of them were currently smoking \citep{3}. These two studies showed that the frequency of smokers was low among the patients with Covid-19. On the other hand, a study conducted by Guan et al. on 1099 patients with Covid-19 shows 227 of them had smoked in the past or current \citep{4}. Therefore, the ratio of smokers to the total population was high in this study. 

The Covid-19 disease may not necessarily be a good sign of whether or not smoking affects patients because anyone can get the virus by being exposed to it. However, the symptoms of this disease are more severe in some people, and usually, these people go to the hospital with signs of a lung infection. Therefore, since the aim is to measure the effect of smoking on Covid-19 patients, this study is only focused on hospitalized patients to examine the rate of smokers.

In this article, we want to find a significant difference between the prevalence rate of smoking cigarettes among people with Covid-19 disease and this rate in the population? In this way, it can be determined whether smoking cigarettes increases or decreases the risk of exacerbating the disease in patients with Covid-19.

\section{Data description}
Data related to 1040 patients with Covid-19 admitted to hospitals in Iran have been collected \citep{5}. These patients were randomly selected from patients admitted to hospitals in Rasht, Tehran, and Bojnord. Of these 1040 patients,  375 of them are female, and 665 of them are male. Also, the age of these people is between 14 and 91 years, and the average age is about 54 years. Unfortunately, 91 of these people lost their lives due to the disease, and 949 recovered. The dispersion distribution of this sample is shown in Table~\ref{tab:table1}.

As Table~\ref{tab:table1} has seen, in this sample of 1040 hospitalized patients, 60 patients smoke cigarettes. Since this study seeks to discover the significant difference in smoking between Covid-19 disease patients and the general population, it is necessary to assess the prevalence of tobacco in the country.

Various studies have been conducted on the prevalence of smoking in the country. For example, Ebadi et al. Surveyed 27883 people and showed that 25.4\% of 18 to 65 years people in Iran smoke \citep{6}. Another study conducted by Farshidi on 1506 people older than 15 in Hormozgan province found that 19.5\% of people living in this province smoke cigarettes \citep{7}.

Also, an extensive survey was conducted by the Health Research Institute of the Islamic Republic of Iran by taking a sample of 30,150 people from all provinces except Qom province, which shows that 18.44\% of Iranians over 18 years of age had a history of smoking and 9.04\% of them are currently smoking cigarette \citep{8}. Furthermore, this survey reported the rate of smoking in more specific groups of people. Its results are shown in Table~\ref{tab:table2} and Table~\ref{tab:table3}. Table~\ref{tab:table2} shows that the smoking percentage in just three provinces because the sample has been collected from hospitals of the three Iranian provinces’ capitals. As this survey is the most comprehensive and up-to-date survey available and officially published, the statistics published in this survey will be the basis for this study.

\begin{table}
	\caption{Data at a glance}
	\centering
	\begin{tabular}{ll}
		\toprule
		Size of Sample   & 1040 patients     \\
		\cmidrule(r){1-2}
		Max age			 & 91 years  		 \\
		\cmidrule(r){1-2}
		Min age		     & 14 years		   	\\
		\cmidrule(r){1-2}
		Mean age	     & 54 years      	\\
		\cmidrule(r){1-2}
		Female			& 375 persons 		\\
		\cmidrule(r){1-2}
		Male			& 665 persons 		\\
		\cmidrule(r){1-2}
		Dead			& 91 persons		\\
		\cmidrule(r){1-2}
		Recovered		& 949 persons		\\
		\cmidrule(r){1-2}
		Smoking cigarette & 60 persons		\\
		\bottomrule
	\end{tabular}
	\label{tab:table1}
\end{table}

\begin{table}
	\caption{Percentage of smoking in different provinces}
	\centering
	\begin{tabular}{llll}
		\toprule
		Province of residence& Male  &Female  &Total    \\
		\midrule
		Tehran 				 & 19.2	 &1.12	  &10.04    \\
		Gilan 				 & 20.98 &0.28	  &10.38    \\
		North Khorasan		 & 8.64	 &0.5	  &4.54     \\
		\bottomrule
	\end{tabular}
	\label{tab:table2}
\end{table}

\begin{table}
	\caption{Percentage of smoking in different ages}
	\centering
	\begin{tabular}{llll}
		\toprule
		Age (years)			 & Male  &Female  &Total    \\
		\midrule
		18-24 				 & 4.44	 &0.04	  &2.24     \\
		25-39 				 & 16.80 &0.56	  &8.76     \\
		40-64				 & 27.14 &1.12	  &14.02    \\
		Over 65				 & 15	 &1.86	  &8.08     \\
		\bottomrule
	\end{tabular}
	\label{tab:table3}
\end{table}

\section{Data analysis}
Based on what was mentioned in the previous section, only 60 out of 1040 critically ill patients with Covid-19 smoke cigarettes. Therefore, the prevalence rate of cigarettes in the sample is equal to 5.77 percent. While this rate, as observed, is equal to 9.04\% in the whole society. We now want to examine whether these data can prove a significant difference between the prevalence of smoking in critically ill Covid-19 patients and society. We know that the distribution of patients with Covid-19 is not equal to the distribution of Iranian society. So we can not compare these two rates directly, and we have to use the rates at different ages, sexes, and locations to find the exact rate of smoking in our data distribution. However, before that, it should be checked what the quality of the sample taken is.

In order to generalize the result obtained from the sample to the community, the sample must be taken randomly. One of the methods to measure the randomness of the sample is to use the randomness runs test \citep{9}. A run means a series of similar observations in a row, and therefore the order of sampling in this test is essential. The number of runs in sampling is crucial, and if it is more or less a limit, the quality of the sampling can be distrusted.
Now we first measure the quality of the sample using the one-sample runs test. For this purpose, IBM SPSS statistical software was used, and the significance level was determined to 5\%. Fortunately, the hypothesis that the sampling is random is not rejected, which means that the data quality is good. Table~\ref{tab:table4} shows the software output. The significant value in this table shows that if we want to reject the assumption that the sample is random, we have a 37.8\% probability of error (type 1 error). Because this value is much higher than the desired level of significance, the hypothesis of the randomness of sampling can not be rejected. Therefore, these data can be used to conclude the entire population of critically ill patients with Covid-19.

\begin{table}
	\caption{Result of One-Sample Runs Test}
	\centering
	\begin{tabular}{l p{5cm} l l p{2.2cm}}
		\toprule
		No.	& Null Hypothesis  &Test  &Sig.   &Decision    \\
		\midrule
		1   & The sequence of values defined by smoke = (0) and (1) is random  & One-Sample Runs Test & 0.378 & Retain the null hypothesis \\
		\midrule
		\multicolumn{5}{l}{Asymptotic significances are displayed. The significance level is 0.050}\\
		\bottomrule
	\end{tabular}
	\label{tab:table4}
\end{table}

After it was shown that the sample was random, it is now possible to examine the difference in smoking prevalence between Covid-19 patients and the population. We have to use statistical hypothesis testing to determine any significant difference between the two populations.
The chi-square test ($\chi ^ 2$), also known as the chi-square or chi-square test, was first proposed by Carl Pearson in the 1890s. He called it the test of goodness of fit. The chi-square test examines the fit of observations and the expected theoretical value when data are classified by only one variable or one dimension \citep{9}. In this test, there is only one variable that is classified into different categories. Also, no more than 20\% of the categories should have an expected value of less than 5. If this is the case, it is recommended that these categories are merged with their neighboring categories, and more significant categories are defined \citep{8}.
To use the chi-square test, we have to compute the expected and observed number of smokers based on the sample data and smoking prevalence rate. Table~\ref{tab:table5} is shown the expected and the observed number of smokers based on their sexes and provinces. Observed numbers calculated based on our collected sample and expected numbers calculated based on Table~\ref{tab:table2} and collected sample.

\begin{table}
	\caption{Expected and the observed number of smokers based on their sexes and provinces}
	\centering
	\begin{tabular}{lllllll}
		\toprule
		\multicolumn{1}{c}{}			&
		\multicolumn{3}{c}{Male}	&
		\multicolumn{3}{c}{Female}		\\
		\cmidrule{2-7}
		Province & Total & Observed & Expected & Total & Observed & Expected \\
		\midrule
		Tehran			& 544 & 51 & 104.45 & 266 & 1 & 2.98	\\
		Gilan			& 83  & 7  & 17.41  & 49  & 0 & 0.14	\\
		North Khorasan  & 38  & 1  & 3.28	& 60  & 0 & 0.3		\\
		\midrule
		Sum				& 665 & 59 & 125.14 & 375 & 1 & 3.42	\\
		\bottomrule
	\end{tabular}
	\label{tab:table5}
\end{table}

Although the observed values in all sexes and provinces are more diminutive than expected values, we cannot use the chi-square test for all data because expected female smokers and North Khorasan males are less than 5. IBM SPSS has been used to determine any significant difference between observed and expected numbers of smokers in other parts of the data. The results is shown in Table~\ref{tab:table6}.

\begin{table}
	\caption{Results of Chi-Square Tests on specific populations}
	\centering
	\begin{tabular}{l p{3cm} p{5cm} p{2.3cm} l p{2.2cm}}
		\toprule
		No.	& Population & Null Hypothesis  &Test  &Sig.   &Decision    \\
		\midrule
		1   & Males living in Tehran Province & The categories of Smoke occur with the specified probabilities.  & One-Sample Chi-Square Test & 0.000 & Reject the null hypothesis \\
		2   & Males living in Gilan Province & The categories of Smoke occur with the specified probabilities.  & One-Sample Chi-Square Test & 0.005 & Reject the null hypothesis \\
		3   & Males living in one of the three provinces & The categories of Smoke occur with the specified probabilities.  & One-Sample Chi-Square Test & 0.000 & Reject the null hypothesis \\
		\midrule
		\multicolumn{5}{l}{Asymptotic significances are displayed. The significance level is 0.050}\\
		\bottomrule
	\end{tabular}
	\label{tab:table6}
\end{table}

The chi-square tests show a significant difference in the frequency of smokers between patients with Covid-19 and the whole population for the populations of males living in Tehran province, males living in Gilan province, and males living in the three provinces. So we can accept that in the test populations, the smoking prevalence of patients with Covid-19 is less than the smoking prevalence in whole people.
Same as Table~\ref{tab:table5}, Table~\ref{tab:table7} is shown the expected and the observed number of smokers based on their sexes and ages. Observed numbers calculated based on our collected sample and expected numbers calculated based on Table~\ref{tab:table3} and collected sample. Although the sample size is 1040, the ages of 10 of them are missed or less than 18. So in Table~\ref{tab:table7}, information of the 1030 sample has been shown.

\begin{table}
	\caption{Expected and the observed number of smokers based on their sexes and ages}
	\centering
	\begin{tabular}{lllllll}
		\toprule
		\multicolumn{1}{c}{}			&
		\multicolumn{3}{c}{Male}	&
		\multicolumn{3}{c}{Female}		\\
		\cmidrule{2-7}
		Age & Total & Observed & Expected & Total & Observed & Expected \\
		\midrule
		18-24			& 12 & 0 & 0.53 & 3 & 0 & 0	\\
		25-39			& 112  & 9  & 18.82  & 51  & 0 & 0.29	\\
		40-64  & 390  & 29  & 105.85	& 210  & 0 & 2.35		\\
		Over 65  & 145  & 20  & 21.75	& 107  & 1 & 1.99		\\
		\midrule
		Sum				& 659 & 58 & 146.95 & 371 & 1 & 4.63	\\
		\bottomrule
	\end{tabular}
	\label{tab:table7}
\end{table}

Although the observed values in all sexes and ages are more diminutive than expected values, we cannot use the chi-square test for all data because expected female smokers and 18 to 24 years old males are less than 5. IBM SPSS has been used to determine any significant difference between observed and expected numbers of smokers in other parts of the data. The results is shown in Table~\ref{tab:table8}.

\begin{table}
	\caption{Results of Chi-Square Tests on specific populations}
	\centering
	\begin{tabular}{l p{3cm} p{5cm} p{2.3cm} l p{2.2cm}}
		\toprule
		No.	& Population & Null Hypothesis  &Test  &Sig.   &Decision    \\
		\midrule
		1   & 25 to 39 years old males & The categories of Smoke occur with the specified probabilities.  & One-Sample Chi-Square Test & 0.013 & Reject the null hypothesis \\
		2   & 40 to 64 years old males & The categories of Smoke occur with the specified probabilities.  & One-Sample Chi-Square Test & 0.000 & Reject the null hypothesis \\
		3   & Over 65 years old males & The categories of Smoke occur with the specified probabilities.  & One-Sample Chi-Square Test & 0.684 & Retain the null hypothesis \\
		\midrule
		\multicolumn{5}{l}{Asymptotic significances are displayed. The significance level is 0.050}\\
		\bottomrule
	\end{tabular}
	\label{tab:table8}
\end{table}

The chi-square tests show a significant difference in the frequency of smokers between patients with Covid-19 and the whole population for the populations of 25 to 39 years old males and 40 to 64 years old males. Nevertheless, for males that are older than 65 years old, we did not find any significant difference between these frequencies, and the null hypothesis was not rejected.
In the end, we want to use all of the data and evaluate the expected smoking prevalence based on the age and sex of the observations. So we use table~\ref{tab:table7} to calculate the expected and observed frequency of smokers according to the distribution of sample data. Table~\ref{tab:table9} shows the expected and observed frequency of smokers, and based on it, the chi-square test results have been shown in Table~\ref{tab:table10}.

\begin{table}
	\caption{The observed and expected frequency of smokers according to the distribution of sample data.}
	\centering
	\begin{tabular}{lll}
		\toprule
		Sample Size   & Observed smokers & Expected smokers     \\
		\midrule
		1030	& 59 & 151.58 		 \\
		\bottomrule
	\end{tabular}
	\label{tab:table9}
\end{table}

\begin{table}[H]
	\caption{Results of Chi-Square Tests on population}
	\centering
	\begin{tabular}{l p{5cm} l l p{2.2cm}}
		\toprule
		No.	& Null Hypothesis  &Test  &Sig.   &Decision    \\
		\midrule
		1   & The categories of Smoke occur with the specified probabilities.  & One-Sample Chi-Square Test & 0.000 & Reject the null hypothesis \\
		\midrule
		\multicolumn{5}{l}{Asymptotic significances are displayed. The significance level is 0.050}\\
		\bottomrule
	\end{tabular}
	\label{tab:table10}
\end{table}

The chi-square tests show a significant difference in the frequency of smokers between patients with Covid-19 and the Iranian population. So we can accept that the smoking prevalence of patients with Covid-19 is less than the smoking prevalence in the population.

\section{Conclusion}
This study examines 1040 patients with Covid-10 admitted to hospitals in Tehran, Rasht, and Bojnord, the capitals of three different Iranian provinces. The one-sample runs test showed that randomness of sampling could not reject (p-value = 0.378), and the result obtained from the sample can be generalized to the population.
Furthermore, The chi-square test showed that the percentage of smokers among hospitalized patients with Covid-19 is significantly lower than that of smokers in the country (p-value < 0.0001). Also, we examined this hypothesis in some sub-populations. Results show a significant difference in the frequency of smokers between patients with Covid-19 and the population of males living in Tehran province (p-value < 0.0001), males living in Gilan province (p-value = 0.005), males living in the three provinces (p-value < 0.0001), 25 to 39 years old males (p-value = 0.013), and 40 to 64 years old males (p-value < 0.0001). Nevertheless, there is no sufficient evidence to reject the null hypothesis for the populations of over 65 years old males.
Although observed female smokers are less than expected female smokers, because of the lack of female patients in our sample (375 out of 1040) and meager rate for Iranian female smokers (only 0.74\% of Iranian women smoke cigarettes \citep{8}), expected female smokers in our samples is very low. So it is not possible to use the chi-square test to determine differences in the female population, and larger sample size is needed.

\section{Data availability statement}
The data that support the findings of this study are openly available in “figshare” at \url{https://doi.org/10.6084/m9.figshare.14853939.v2}.

\bibliographystyle{unsrtnat}
\bibliography{references}

\end{document}